\documentclass[10pt,prl,aps,twocolumn,superscriptaddress]{revtex4-1}
\usepackage{graphicx}
\usepackage{amssymb}
\usepackage{amsmath}
\usepackage{bm}

\newcommand{\bc}{\begin{center}}
\newcommand{\ec}{\end{center}}
\def\ba#1{\begin{array}{#1}\displaystyle}
\newcommand{\ea}{\end{array}}

\newcommand{\beq}{\begin{equation}}
\newcommand{\eeq}{\end{equation}}
\newcommand{\beqa}{\begin{eqnarray}}
\newcommand{\eeqa}{\end{eqnarray}}

\newcommand{\n}{\nonumber\\}
\newcommand{\bi}{\begin{itemize}}
\newcommand{\ei}{\end{itemize}}

\def\lt#1{\left#1}
\def\rt#1{\right#1}
\def\t#1{\tilde{#1}}
\def\h#1{\hat{#1}}

\def\frc#1#2{\frac{#1}{#2}}

\newcommand{\bra}{\langle}
\newcommand{\ket}{\rangle}
\newcommand{\Z}{{\mathbb{Z}}}

\newcommand{\C}{{\mathbb{C}}}
\newcommand{\Tr}{{\rm Tr}}
\newcommand{\TT}{{\cal T}}

\newcommand{\ep}{\epsilon}

\newcommand{\ri}{{\rm i}}

\begin{document}

\title{Entanglement Content of Quasi-Particle Excitations}

\author{Olalla A. Castro-Alvaredo}
\affiliation
{Department of Mathematics, City, University of London, 10 Northampton Square, EC1V 0HB, London, U.K.}
\author{Cecilia De Fazio}
\affiliation
{Dipartimento di Fisica e Astronomia, Universit\`a di Bologna, Via Irnerio 46, I-40126 Bologna, Italy}
\author{Benjamin Doyon}
\affiliation
{Department of Mathematics, King's College London, Strand WC2R 2LS, London, U.K.}
\author{Istv\'an M. Sz\'ecs\'enyi}
\affiliation
{Department of Mathematics, City, University of London, 10 Northampton Square, EC1V 0HB, London, U.K.}

\date{\today}

\begin{abstract}
We investigate the quantum entanglement content of quasi-particle excitations in extended many-body systems. We show that such excitations give an additive contribution to the bi-partite von Neumann and R\'enyi entanglement entropies that takes a simple, universal form. It is largely independent of the momenta and masses of the excitations, and of the geometry, dimension and connectedness of the entanglement region. The result has a natural quantum information theoretic interpretation as the entanglement of a state where each quasi-particle is associated with two qubits representing their presence within and without the entanglement region, taking into account quantum (in)distinguishability. This applies to any excited state composed of finite numbers of quasi-particles with finite De Broglie wavelengths or finite intrinsic correlation length. We derive this result analytically in one-dimensional massive bosonic and fermionic free field theories and for simple setups in higher dimensions. We provide numerical evidence for the harmonic chain and the two-dimensional harmonic lattice in all regimes where excitations have quasi-particle properties. Finally, we provide supporting calculations for integrable spin chain models and other situations without particle production. Our results point to new possibilities for creating entangled states using many-body quantum systems.

\end{abstract}

\maketitle

{\em Introduction.---} Measures of entanglement, such as the entanglement entropy (EE) \cite{bennet} and entanglement negativity \cite{VW, ZHSL, ple, erratumple, Eisert, Eisert2}, have attracted much attention in recent years, both theoretically \cite{EERev1,specialissue, EERev2} and experimentally \cite{Greiner1,Greiner2}. Quantum entanglement encodes correlations between degrees of freedom associated to independent factors of the Hilbert space, and as such, it separates quantum correlations from the particularities of observables.  As a consequence, the entanglement in extended systems encodes, in a natural fashion, universal properties of the state. For instance, at criticality, the entanglement of ground states provides an efficient measure of universal properties of quantum phase transitions, such as the (effective) central change of the corresponding conformal field theory (CFT) and the primary operator content \cite{CallanW94,HolzheyLW94,latorre1,Latorre2,Calabrese:2004eu,Calabrese:2005in,disco1,negativity1,negativity2,BCDLR}. 
Near criticality, it is universally controlled by the masses of excitations \cite{entropy,next,ourneg}. 
In states that are highly excited, with finite energy densities, the entanglement is known to give rise to local thermalisation effects: this is at the heart of the eigenstate thermalisation hypothesis \cite{ETH1,ETH2,ETH3,ETH4,ETH5}, as the large entanglement between local degrees of freedom and the rest of the system effectively generates a Gibbs ensemble (in the case of integrable systems, a generalized Gibbs ensemble). The entanglement effects of a finite number of excitations are less known. Some results are available in critical systems: using the methods of Holzhey, Larsen and Wilczek \cite{HolzheyLW94}, combining a geometric description with Riemann uniformization techniques in CFT it was shown in  \cite{german1,german2}  that certain excitations, with energies tending to zero in the large volume limit, correct the ground state entanglement by power laws in the ratio of length scales. Various few-particle states have also been studied in special cases of integrable spin chains \cite{Vincenzo,ln2,SS,Berko,Vincenzo2}.

In this paper we propose a universal formula, with a simple quantum information theoretic interpretation, for the entanglement content of states with well-defined quasi-particle excitations. For this purpose, we study a variety of extended systems of different dimensions. We consider the von Neumann and R\'enyi EEs: these are measures of the amount of quantum entanglement, in a pure quantum state, between the degrees of
freedom associated to two sets of independent observables whose union is complete on the Hilbert space. We use the setup where the Hilbert space is factorised as ${\cal H}_A\otimes {\cal H}_B$, according to two complementary spatial regions $A$ and $B$, of typical length scales $\ell_A$ and $\ell_B$, respectively (for dimensions higher than one, we can think of these as the diameters of the regions under consideration). The regions can be of generic geometry and connectedness. Quasi-particle excitations arise naturally in massive quantum field theory (QFT), where irreducible representations of the Poincar\'e group are identified with relativistic particles. We first consider excited states of the massive free real boson and free Majorana fermion models, formed of finite numbers $k$ of particles, at various momenta. We use techniques based on form factors of branch point twist fields in 1+1 dimensions as introduced in \cite{entropy}, and dimensional reduction methods \cite{KG} to access simple entanglement regions in higher dimensions. More generally, excitations in the harmonic chain and in higher-dimensional harmonic lattices can be interpreted as being composed of quasi-particles whenever the correlation length $\xi$ is small enough, $\xi\ll{\rm min}(\ell_A,\ell_B)$, or the maximal De Broglie wavelength of the particles $\zeta = {\rm max}\{2\pi/|\vec p_i|\}$, where $\vec p_i$ are the particles' momenta, is small enough, $\zeta\ll {\rm min}(\ell_A,\ell_B)$. This includes the large-momenta region of the chain's or lattice's CFT regime (beyond the applicability of the results of \cite{german1,german2}), as well as the non-universal regime, beyond QFT and CFT. We then study the harmonic chain and two-dimensional lattice numerically. Quasi-particles also arise naturally in Bethe ansatz integrable models, where the conditions $\xi\ll{\rm min}(\ell_A,\ell_B)$ and $\zeta\ll {\rm min}(\ell_A,\ell_B)$ can be given precise meanings. We study few-particle excitations in generic states of the Bethe form. In this case, the calculation is elementary, and complements some of the calculations done in \cite{Vincenzo}. Zamolodchikov's ideas underpinning the thermodynamic Bethe ansatz \cite{tba1,tba2} in fact suggest that the same calculation should also give the correct answer in massive integrable QFT, thus including the effects of interactions. In all cases, we find a universal result in the limit ${\rm min}(\xi,\zeta)\ll{\rm min}(\ell_A,\ell_B)$, independent of the model studied, of the connectedness or shape of the entanglement region, and of the dimension. The result extends the ``semiclassical" form discussed in the context of spin chains in \cite{Vincenzo}. It has a very natural qubit interpretation where qubits representing the particles are entangled according to the particles' distribution in space, taking into account quantum indistinguishability in the bosonic case.  The qubit interpretation indicates that quasi-particle excitations in many-body models may provide a simple way of generating entangled states with easily adjustable parameters.

\begin{figure}[h!]
\begin{center} 
 \includegraphics[width=4.4cm]{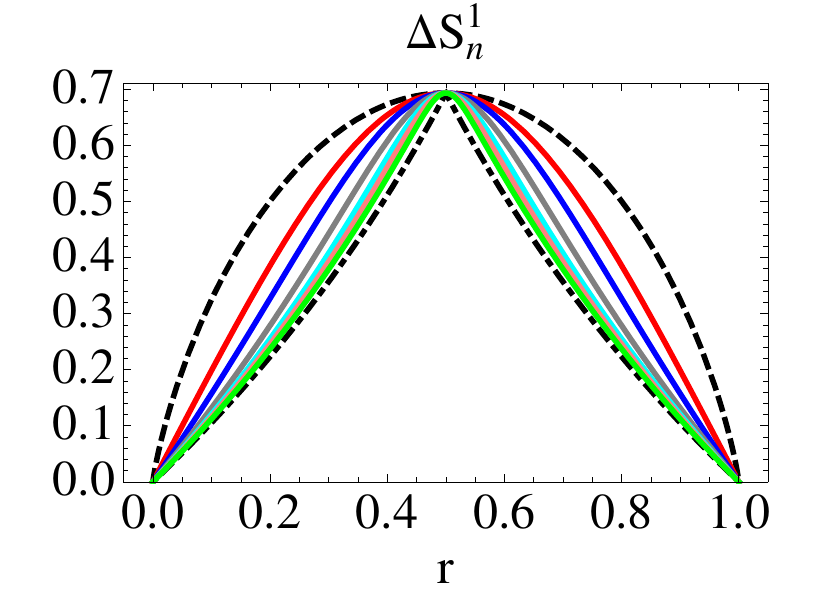} 
 \hspace{-0.5cm}
   \includegraphics[width=4.4cm]{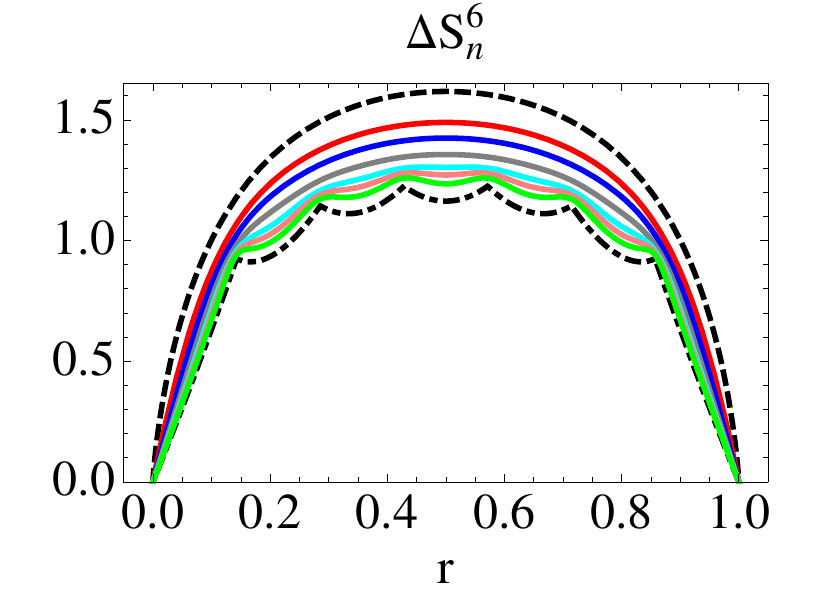} 
 \end{center} 
 \caption{The functions (\ref{result1}) and (\ref{3}) for $k=1, 6$ and  $n=2,3,5,8, 11, 17$ and the limits $n \rightarrow 1$ (von Neumann) and $n\rightarrow \infty$ (single-copy). The outer-most curve is the von Neumann entropy and the inner-most curve is the single-copy entropy.} 
 \label{cuspy} 
  \end{figure}

{\em Results.---}  Consider a bi-partition of a system  $C=A \cup B$ in state $|\Psi\ket$ composed of a number $k$ of quasi-particles. In infinite volume, the notion of quasi-particles is very natural via the theory of scattering states \cite{poles,qft}, and one expects that in finite but large volumes there are corresponding excited states, unambiguously defined up to exponentially decaying corrections in the volume. We define the reduced density matrix associated to subsystem $A$ as $\rho_A=\Tr_{B}(|\Psi \rangle \langle \Psi|)$. The R\'enyi EE is the R\'enyi entropy of this reduced density matrix,  
\beq 
 S_n^\Psi(A,B)=\frac{\log \Tr\rho_A^n }{1-n}.
 \label{renyi}
 \eeq 
From (\ref{renyi}) we may compute the von Neumann EE as $S^\Psi_1 (A,B):=\lim_{n\rightarrow 1} S_n^\Psi(A,B)$ and 
the so-called single-copy entropy \cite{EC, PZ, DD} as $S_\infty^\Psi(A,B):= \lim_{n\rightarrow \infty} S_n^\Psi(A,B)$.
For large system size and fixed entanglement region, one expects the entanglement entropies to tend to those of the ground state. We therefore concentrate on the nontrivial limit where both the full system $C$ and the entanglement region $A$ are large, scaled simultaneously, $A\mapsto \lambda A$ and $B\mapsto \lambda B$. Let $r = {\rm Vol}_d\,(A) / {\rm Vol}_d\,(C)$ be the ratio of the $d$-dimensional hypervolume of the region to that of the system. We compute the difference $\Delta{S}_n^\Psi(A,B) = S_n^\Psi(A,B) - S_n^0(A,B)$ between the R\'enyi entropy in the excited state $|\Psi\ket$ and in the ground (vacuum) state $|0\ket$, in this limit,
\beq
	\Delta{S}_n^\Psi(r) := \lim_{\lambda\to\infty} \Delta{S}_n^\Psi(\lambda A,\lambda B).
	\label{limit}
\eeq
This is the contribution of the excitations to the entanglement, or ``excess entanglement" as named in \cite{german1,german2}.

We find that for a wide variety of quantum systems, the results depend only on the proportion $r$ of the system's volume occupied by the entanglement region, and are largely independent of the momenta of the quasi-particles. Suppose the state is formed of $k$ particles of equal momenta. Denoting $\Delta S_n^\Psi(r) = \Delta S_n^k(r)$, we find
\beq
\Delta{S}_n^{k}(r)=\frac{\log
\sum\limits_{q=0}^k f_q^k(r)^n
}{1-n},\ 
\Delta{S}^{k}_1(r)=-{ \sum\limits_{q=0}^k f_q^k(r)\log f_q^k(r)}
\label{result1}
\eeq
with $f_q^k(r) =\left(\begin{array}{c}
k\\
q
\end{array}\right) r^{q} (1-r)^{k-q}$. For a state composed of $k$ particles divided into groups of $k_i$ particles of equal momenta $\vec p_i$, with $i=1,2,\ldots$ and $\sum_i k_i = k$, we denote $\Delta S_n^\Psi(r) = \Delta S_n^{k_1,k_2,\ldots}(r)$ and have
\beq\label{result2}
	\Delta S_n^{k_1,k_2,\ldots}(r) = \sum_i \Delta{S}_n^{k_i}(r).
\eeq
In particular, for $k$ particles of distinct momenta the result is $k$ times that for a single particle, which is
\beqa
\Delta{S}_n^{1}(r)&=&\frac{\log(r^{n}+(1-r)^n)}{1-n},\n
\Delta{S}^{1}_1(r)&=&-r \log r - (1-r)\log(1-r).
\label{vn}
\eeqa

We observe that in all cases, the entanglement is maximal at $r=1/2$. For $k$ distinct-momenta particles, the maximum is $k\log 2$, while when some particles have coinciding momenta, the maximal value is smaller. Interestingly, single-copy entropies present non-analytic features. For distinct momenta, we have
\beq
\Delta{S}_\infty^{1}(r)=\left\{\begin{array}{ll} -\log(1-r) & \mathrm{for}\quad 0\leq r <\frac{1}{2}\\
 -\log r & \mathrm{for}\quad \frac{1}{2}\leq r \leq 1.
 \end{array} \right.
 \label{single}
\eeq
Again, the result is just multiplied by $k$ for a state consisting of $k$ distinct-momentum particles. 
For equal momenta it is a function which is non-differentiable at $k$  points in the interval $r\in(0,1)$ (generalizing (\ref{single})).  The positions of these cusps are given by the values
\beq
r=\frac{1+q}{1+k} \quad \mathrm{for} \quad q=0,\ldots,k-1,
\eeq
and the single copy entropy is given by
\beq
 \Delta{S}_\infty^{k}(r)=
 -\log f_q^k(r)  \quad \mathrm{for}\quad \frac{q}{1+k}\leq r < \frac{1+q}{1+k} 
\label{3}
\eeq
and $q=0, \dots, k$. 

The results take their full meaning under a quantum information theoretic interpretation that combines a ``semiclassical'' picture of particles with quantum indistinguishability. Consider a bi-partite Hilbert space ${\cal H} = {\cal H}_{\rm int}\otimes {\cal H}_{\rm ext}$. Each factor ${\cal H}_{\rm int}\simeq {\cal H}_{\rm ext}$ is a tensor product $\otimes_i {\cal H}^{k_i}$ of Hilbert spaces ${\cal H}^{k_i}\simeq \C^{k_i+1}$ for $k_i$ indistinguishable qubits, with, as above, $\sum_i k_i = k$. We associate ${\cal H}_{\rm int}$ with the interior of the region $A$ and ${\cal H}_{\rm ext}$ with its exterior, and we identify the qubit state $1$ with the presence of a particle and $0$ with its absence. We construct the state $|\Psi_{\rm qb}\ket\in {\cal H}$ under the picture according to which equal-momenta particles are indistinguishable, and a particle can lie anywhere in the full volume of the system with flat probability. That is, any given particle has probability $r$ of lying within $A$, and $1-r$ of lying outside of it. We make a linear combination of qubit states following this picture, with coefficients that are square roots of the total probability of a given qubit configuration, taking proper care of (in)distinguishability. For instance, for a single particle,
\beq
	|\Psi_{\rm qb}\ket= \sqrt{r} \;|1\ket \otimes |0\ket + \sqrt{1-r}\;|0\ket \otimes |1\ket
	\label{2qubit}
\eeq
as either the particle is in the region, with probability $r$, or outside of it, with probability $1-r$. If two particles of coinciding momenta are present, then we have
\beqa
	|\Psi_{\rm qb}\ket &=& \sqrt{r^2}\; |2\ket \otimes |0\ket + \sqrt{2r(1-r)}\;|1\ket \otimes |1\ket\n
	&& +\, \sqrt{(1-r)^2}\;|0\ket \otimes |2\ket
\eeqa
as either the two particles are in the region, with probability $r^{2}$, or one is in the region and one outside of it (no matter which one), with probability $2r(1-r)$, or both are outside the region, with probability $(1-r)^2$. For two particles of different momenta,
\beqa
	|\Psi_{\rm qb}\ket &= &\sqrt{r^{2}}\; |11\ket \otimes |00\ket + \sqrt{(1-r)^2}\;|00\ket \otimes |11\ket\n && +\, \sqrt{r(1-r)}\;(|10\ket \otimes |01\ket + |01\ket \otimes |10\ket)
\eeqa
counting the various ways two distinct particles can be distributed inside or outside the region. Higher-particle states can be constructed similarly. The results stated above are then equivalent to the identification $\Delta S_n^\Psi(r) = \Delta S_n^{\Psi_{\rm qb}}(r) $, where
\beq\label{qubit}
	 \Delta S_n^{\Psi_{\rm qb}}(r) = \frc{\log\Tr\rho_{{\rm int}}^n}{1-n},\quad
	\rho_{{\rm int}} = \Tr_{{\rm ext}} |\Psi_{\rm qb}\ket\bra\Psi_{\rm qb}|.
\eeq

{\em Methods.---} In general, the quantity $\Delta S_n^{\Psi}$ can be computed using the replica method \cite{CallanW94,HolzheyLW94}. In this context, one evaluates traces of powers of the reduced density matrix $\rho_A$. After reinterpretation of such traces, this boils down to ratios of expectation values of a twist operator, acting on a replica model composed of $n$ independent copies of the original theory. The operator $\mathbb{T}(A,B)$ acts as a cyclic permutation of the copies $i\mapsto i+1\ {\rm mod}\ n$ on ${\cal H}_A$, and as the identity on ${\cal H}_B$, and (\ref{limit}) is expressed as
\beq
 \Delta{S}_n^\Psi(r)
= \lim_{\lambda\to \infty}
\frac{1}{1-n}{\log\left[\frac{{}_n\bra \Psi|\mathbb{T}(\lambda A,\lambda B)|\Psi \ket_n}{{}_n\bra 0|\mathbb{T}(\lambda A,\lambda B)|0 \ket_n}\right]},
\label{gen}
\eeq
where $|0\ket_n$ is the vacuum state. Both $|0\ket_n$ and the state $|\Psi\ket_n$ have the structure
\beq
|\Psi\ket_n=  |\Psi\ket^1 \otimes |\Psi\ket^2 \otimes \cdots \otimes |\Psi\ket^n.
\label{replicastate}
\eeq
Here $|\Psi\ket^i \simeq |\Psi\ket$ is the $k$-particle excited state of interest, implemented in the $i$th copy. 

In one dimension, $A$ is in general a union of segments. Then, $\mathbb{T}(A,B)$ is expressed as a product of branch-point twist fields \cite{entropy}, supported on the boundary points of these segments. Branch point twist fields are twist fields associated to the cyclic permutation, a symmetry of the replica model. Let us consider the case $A = [0,\ell]$ in a system of length $L$. Then $\mathbb{T}(A,B) = {\mathcal T}(0)\t{\mathcal T}(\ell)$, where $\mathcal{T}$ is the branch point twist field and $\tilde{\TT}$ is its hermitian conjugate. Expression (\ref{gen}) may be used by expanding two-point functions of branch-point twist fields in the basis $|\Phi\ket$ of quasi-particles,
\beq\label{ffexpansion}
	{}_n\bra \Psi|\mathcal{T}(0)\tilde{\mathcal T}(\ell)|\Psi \ket_{n} = \sum_{\Phi} e^{-\ri P_\Phi \ell}
	|{}_n\bra \Psi|\mathcal{T}(0)|\Phi\ket|^2
\eeq
where $P_\Phi$ are the momentum eigenvalues (in finite volume, they are quantised, and the set of states is discrete).

Using (\ref{gen}) with (\ref{ffexpansion}) in integrable 1+1-dimensional QFT presents however a number of challenges. Via an extension of the form factor program \cite{KW,Smirnovbook}, matrix elements of branch-point twist fields in infinite volume are known exactly \cite{entropy}, and have been used successfully in the vacuum. But they cannot be used in order to evaluate the limit $L\to\infty$ in (\ref{gen}), as in excited states, divergencies occur in the expansion (\ref{ffexpansion}) whenever momenta of intermediate particles in $|\Phi\ket$ coincide with momenta of particles in the state $|\Psi\ket_n$. One must first evaluate finite-volume matrix elements, re-sum the series (\ref{ffexpansion}), and then take the limit. Finite-volume matrix elements of ordinary local fields in integrable QFT have been studied recently \cite{PT1,PT2}. They are simply related to infinite-volume matrix elements up to exponentially decaying terms in $L$. They are evaluated at momenta that are quantised according to the Bethe-Yang equations based on the two-particle scattering matrix of the integrable model. But for twist fields, the theory has not been developed yet. In particular, the twist properties affect the quantisation condition of the individual momenta of the quasi-particles. We have solved these problems for the massive free real boson and the massive free Majorana fermion in periodic space. By performing the summation over intermediate states at large $L$, noting that the so-called ``kinematic singularities" of infinite-volume matrix elements provide the leading contribution, we have derived the full results (\ref{result1}) and (\ref{result2}). The details are technical, and presented in a separate paper \cite{ussoon}.

\begin{figure}[h!]
 \includegraphics[width=4.3cm]{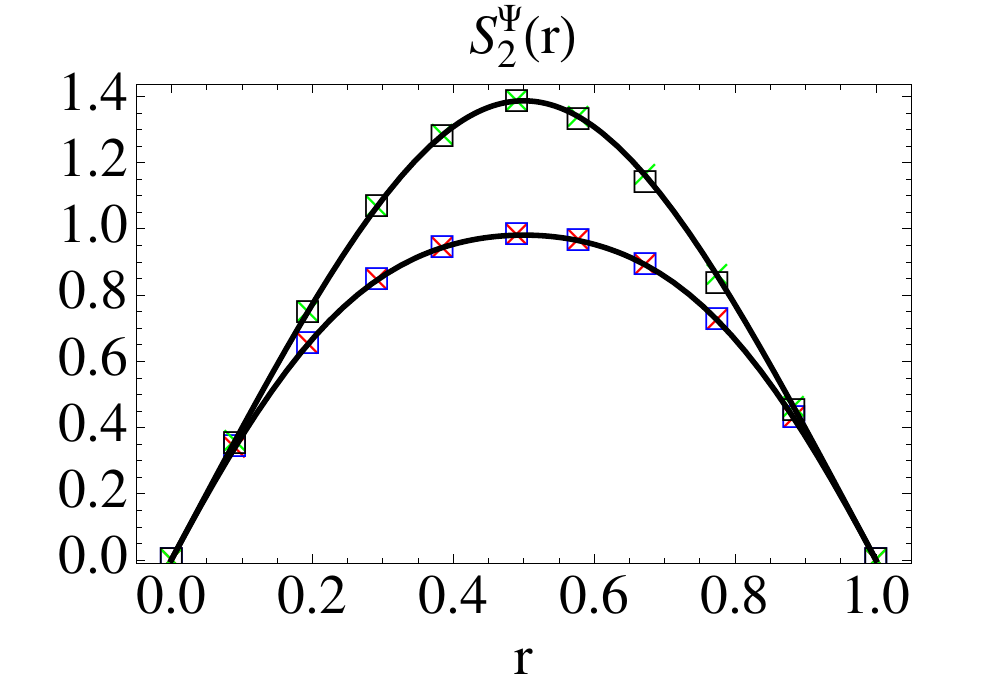} 
 \hspace{-0.2cm}
  \includegraphics[width=4.3cm]{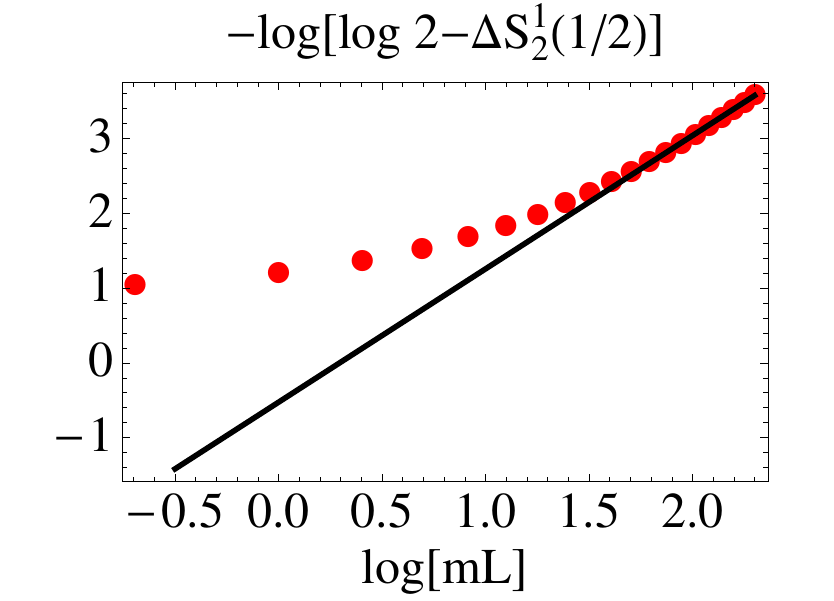} 
\caption{Numerical results for the case $n=2$ on the toric lattice $[0,L]^2$ with $L=50$ and lattice spacing $\Delta x=1$. Left. Two-particle states $|\Psi\ket = |\vec p_1,\vec p_2\ket$. Squares are for mass $m=1$ and small momenta, crosses are for mass $m=0.001$ and large momenta. The upper curve is formula (\ref{result2}) for $\Delta S_2^{1,1}$, with numerical results for distinct momenta $\vec p_1= (0,0),\, \vec p_2=(0.26,0) = (4\pi / L,0)$ (squares) and $\vec p_1=(2.51,1.26)= (40 \pi/L,20 \pi/L),\,\vec p_2=(3.14,0)= (50 \pi/L,0)$ (crosses). The lower curve is formula (\ref{result1}) for $\Delta S_2^{2}$, with numerical results for equal momenta $\vec p_1=\vec p_2=(0.13,0)=(2\pi / L,0)$ (squares) and $\vec p_1=\vec p_2=(2.51,1.26)$ (crosses). Right. Approach of $\Delta S_2^1(1/2)$ to the analytical value $\log 2$ for the one-particle state $|\Psi\ket = |\vec p\ket$ with $\vec p=(0,0)$ as a function of $L$. This shows a linear approach for large $mL$. The solid line is the fit $0.527 - 1.783 \log (mL)$ on the last 8 data points ($mL\in[6.5,10]$).}
\label{numerics}
\end{figure}

The qubit interpretation presented earlier suggests that our results need not be restricted to one-dimensional QFT. To test this claim, we performed a numerical evaluation of the quantity $\Delta S_n^\Psi(r)$ in the harmonic chain and the two-dimensional harmonic lattice. We present some of our results in FIG.~2 and in the supplementary material (SM). 
We used wave-functional methods in order to represent the state $|\Psi\ket$. In the finite-volume Klein-Gordon theory, the vacuum wave functional takes the form
\beq
	\hspace{-0.1cm} \bra \varphi|0\ket \propto \exp\lt[-\frc12 \int_{C\times C} {\rm d}^d x\,{\rm d}^d y\,K(\vec x-\vec y)\varphi(\vec x)\varphi(\vec y)\rt]
\eeq
where $K(\vec x) = \sum_{\vec p}{\rm Vol}_d\,(C)^{-1} E_{\vec p} \,e^{\ri \vec p\cdot \vec x}$. Excited state wave functionals are obtained by applying the operator
\beq\label{creation}
	A^\dag(\vec p) = \frc{\int_C {\rm d}^d x \,e^{\ri \vec p\cdot \vec x}\lt(E_{\vec p}\varphi(\vec x)-\ri\varpi(\vec x)\rt)}{\sqrt{2E_{\vec p}\,{\rm Vol}_{d}\,(C)}},\  [A_{\vec p},A^\dag_{\vec q}] = \delta_{{\vec p},{\vec q}},
\eeq
(where $E_{\vec p} = \sqrt{\vec p^2 + m^2}$) with the representation of the canonical momentum $\varpi(\vec x) = -\ri \delta / \delta\varphi(\vec x)$ satisfying $[\varphi(\vec x),\varpi(\vec y)] = \ri \delta( \vec x-\vec y)$. This generates factors which are polynomial functionals of $\varphi(\vec x)$. The operator $\mathbb{T}(A,B)$ is easily implemented on the space of field configurations. The ratio (\ref{gen}) then becomes a Gaussian average of polynomial functionals of the fields. Discretising space to a finite lattice spacing $\Delta x$ modifies the dispersion relation to $E_{\vec p}^2 = m^2 + 4\sum_{i=1}^d \sin^2 (p_i\Delta x/2)/(\Delta x)^2$. Numerical results in the one-dimensional case are discussed in more detail in \cite{ussoon}, where both QFT and non-universal parameter regimes are seen to agree with our predictions, for connected and disconnected regions. We concentrate here on the two-dimensional periodic square lattice on $C = [0,L]^2$. We choose a set of subregions $A = [0,\ell]^2$ for values of $r=\ell^2/L^2$ ranging between 0 and 1. In order to establish the validity of the requirements on the correlation length $\xi$ and the De Broglie wavelength $\zeta$, we explore two distinct regimes: that of small $\xi$ but large $\zeta$, and that of small $\zeta$ but large $\xi$, in both cases looking at two-particle states with equal and with distinct rapidities. We find excellent agreement with formulae (\ref{result2}) for $\Delta S_n^{1,1}$ and (\ref{result1}) for $\Delta S_n^{2}$, respectively, see FIG.~\ref{numerics}. Note that the configuration we have chosen is not symmetric: regions $A$ and $B$ have different shapes. Nevertheless, the symmetry $r\mapsto 1-r$ is correctly recovered in the regime of validity of formulae (\ref{result1}) and (\ref{result2}). We have explored other shapes of the region $A$, obtaining similar accuracy, and have analysed regimes where both $\xi$ and $\zeta$ are small, finding even greater accuracy. We have also analysed the breaking of formulae (\ref{result1}) and (\ref{result2}) away from their regime of validity. The approach to the maximum $\log 2$ in the case of a single particle with $r=1/2$ (this maximal value is supported by general arguments \cite{ln2}) is shown in FIG.~\ref{numerics}, where the correlation length is varied; we observe an algebraic approach at large $mL$.

For particular choices of the region $A$, it is possible to show analytically the results (\ref{result1}--\ref{3}) in the massive Klein-Gordon and Majorana models of any higher dimension, by dimensional reduction \cite{KG}. Consider the slab-like regions $A = [0,\ell]\times C_\perp$ in $C=[0,L]\times C_\perp$ where $C_\perp$ is some $d-1$-dimensional space. From the $d$-dimensional Klein-Gordon fields $\varphi(\vec x,t)$ and  $\varpi(\vec x,t)$ construct
\beq
	\h\varphi(x_1,t) = \frc{ \int_{C_\perp} {\rm d}^{d-1}x_\perp \varphi(x_1,\vec x_\perp,t)}{\sqrt{{\rm Vol}_{d-1}\,(C_\perp)}}
\eeq
and similarly for $\h\varpi(x_1,t)$ in terms of $\varpi(\vec x,t)$. One observes that $\h\varphi(x,t)$ and $\h\varpi(x,t)$ are canonically normalised one-dimensional Klein-Gordon fields, and that the dimensional reduction map preserves the vacuum \cite{KG}. It also preserves the many-particle states when all momenta point in the $x_1$ direction: with $\vec p = (p,0,\ldots,0)$ the expression (\ref{creation}) gives $A^\dag(\vec p) = \h A^\dag(p)$. Therefore, the quantity ${}_n\bra \Psi|\mathbb{T}(\lambda A,\lambda B)|\Psi \ket_n$ in $d$ dimensions, is proportional to ${}_n\bra \h\Psi|\mathcal{T}(0)\t{\mathcal{T}}(\ell)|\h\Psi \ket_n$ in 1 dimension. The singularity as $\ell\to0$ is dimension dependent, but in the ratio (\ref{gen}), this cancels out, and there is exact equality. See \cite{ussoon} for more details. This analysis extends to other quasi-one-dimensional configurations.

Finally, we establish that our results hold beyond free theories. We analyse the quantity $\Delta S_n^\Psi(r)$ in interacting states of the Bethe ansatz form. Previous analyses exist \cite{Vincenzo,Vincenzo2}, which however concentrated on less universal regimes. In the ferromagnetic Heisenberg chain, two-particle states with respect to the ferromagnetic vacuum have the simple form $\sum_{x,y\in\Z} e^{\ri px + \ri qy} S_{{\rm sgn}(x-y)}(p,q) |\uparrow\cdots \downarrow_x\cdots \uparrow\cdots \downarrow_y\cdots\uparrow\ket$, where $S_\ep(p,q)$ is the Bethe ansatz scattering matrix. As in the context of the thermodynamic Bethe ansatz formalism of integrable QFT \cite{tba1,tba2}, for the purpose of evaluating large-distance quantities these are abstract states representing two-particle asymptotic states, with $S_\ep(p,q)$ the two-body scattering matrix of the field theory. States of the Bethe ansatz form thus are expected to provide large-distance results of great generality in integrable models. We have analysed such one- and two-particle states, and found that formulae (\ref{result1}) and (\ref{result2}) hold, see the SM. There is no need to fix the momenta via the Bethe ansatz; with equal momenta $S_n^2(r)$ is indeed reproduced, extending previous results. Bound states of the Heisenberg chains (Bethe strings) have been studied \cite{Vincenzo}; these have an intrinsic length scale $\xi$ (inversely proportional to the bounding energy), and one can see that in the regimes discussed above, $S_n^1(r)$ is indeed reproduced.


{\em Discussion.---} It is remarkable that the entanglement of a wide variety of many-body quantum systems admits such a simple and universal ``qubit" interpretation.
This combines a semiclassical picture of localised particles controlled by correlation lengths and De Broglie wavelengths, with the quantum effect of (in)distinguishability. The applicability of equations (\ref{result1}--\ref{3}) to higher dimensions is particularly significant, showing that a large amount of geometric information is irrelevant. Their application to QFT is also interesting: QFT locality is formally based on the vanishing of space-like commutation relations, not on particles, yet our results show how quantum entanglement clearly ``sees" localised particles. The results hold beyond the QFT regime, as we checked in quantum harmonic lattices and in Bethe ansatz excitations of quantum spin chains. Going beyond integrability, the results are fully expected to hold when no particle production occurs, for instance in QFT one-particle states, and two-particle states below the particle production threshold. In fact, any one- and two-particle excitations of Bethe-ansatz form will have EE described by  (\ref{result1})-(\ref{3}), such as in spin-preserving quantum chains, integrable or not. The relation (\ref{qubit}) also suggests that quasi-particle excitations in extended systems of any dimension can be used to create simple entangled states with controllable entanglement, where the control parameter is the region-to-system volume ratio $r$. It would be interesting to investigate the possible applications of such a result in the area of quantum information.



\paragraph{Acknowledgements:} We are grateful to EPSRC for funding through the standard proposal ``Entanglement Measures, Twist Fields, and Partition Functions in Quantum Field Theory" under reference numbers EP/P006108/1 and EP/P006132/1. We would also like to thank Vincenzo Alba for discussions and for bringing reference \cite{Vincenzo} to our attention.

\onecolumngrid
\appendix
\section{Supplementary Material }
\section{R\'enyi Entropy of one- and two-Magnon States in Gapped Quantum Spin Chains}
In this section  we present a derivation of the nth R\'enyi entropy for a one-magnon state and the 2nd R\'enyi entropy of a two-magnon  state in a generic gapped quantum spin-$\frac{1}{2}$ chain of length $L$. There is some overlap with calculations presented in the appendices of \cite{Vincenzo} but the focus is slightly different here. An obvious example would be the XXZ model in the gapped regime but other models can also be included, as long as their excited states can be represented as 
\beq
|\Psi\ket_1=\frac{1}{\sqrt{L}}\sum_{j=1}^L e^{i p j} |j\ket,
\label{one}
\eeq
and
\beq
|\Psi\ket_2= \frac{1}{\sqrt{N}}\sum_{j_1,j_2=1}^L S_{j_1 j_2} e^{i p_1 j_1+ip_2 j_2} |j_1 j_2\ket,
\label{two}
\eeq
where 
\beq
S_{j_1 j_2}=(S_{j_1 j_2}^*)^{-1}=\left\{\begin{array}{cc}
e^{i\varphi} & \mathrm{for}\quad j_1>j_2\\
1& \mathrm{for} \quad j_1<j_2\\
0 & \mathrm{for} \quad j_1=j_2
\end{array}\right.
\label{fases}
\eeq
for a two-particle excited state. The states $|j\ket$ and $|j_1 j_2\ket$ represent tensor product states where all spins are up, except the spin at position $j$ or the spins at positions $j_1, j_2$, respectively. The normalization $N$ of the two-particle excited state will be discussed below. In our computations we will consider a bi-partition such that
\beq
\mathrm{spins} \,\, \{1,\ldots,\ell\}\in A \quad \mathrm{and} \quad \mathrm{spins} \,\, \{\ell+1, \ldots,L\}\in B
\eeq
The choice of the spins is relevant in the computations below as it determines the value of the scattering phases (\ref{fases}). However, it is easy to show that this choice is not essential. That is, the results would be identical in the regions $A$ and $B$ are not simply-connected. Finally we note that we will use the notation
\beq
r:=\frac{\ell}{L}
\eeq
to denote the dimensionless ratio of length scales in the problem. 
\subsection{One-Magnon State}
Consider the state  (\ref{one}). The basic states are normalized as $\bra j_1| j_2\ket=\delta_{j_1 j_2}$ and therefore the state above has norm 1.  We would like to compute the R\'enyi entropy of region $A$ for the state $|\Psi\ket$. This computation is actually very simple and the results have been presented for instance in \cite{Berko}. Let us first compute the reduced density matrix
\beq
\rho_A=\mathrm{Tr}_B (|\Psi\ket_1 {}_1\bra \Psi| ).
\eeq
This is
\beqa
\rho_A&=&\frac{1}{L} \mathrm{Tr}_B \left(\sum_{j,k=1}^L e^{ip(j-k)}|j\ket \bra k| \right)\nonumber\\
&=&\frac{1}{L} \sum_{j,k\in A}e^{ip(j-k)}|j\ket \bra k|  \mathrm{Tr}_B\left( |0\ket  \bra 0| \right)+\frac{1}{L}  |0\ket \bra 0| \sum_{j,k\in B}e^{ip(j-k)}\mathrm{Tr}_B\left(|j\ket \bra k|\right)\nonumber\\
&& +\frac{1}{L}\sum_{j\in A} e^{ip j}|j\ket \bra 0|  \sum_{k\in B} e^{-ipk} \mathrm{Tr}_B\left(|0\ket \bra k|\right)+\frac{1}{L}\sum_{k\in A}e^{-i p k}|0\ket \bra k|  \sum_{j\in B} e^{ipj}\mathrm{Tr}_B\left(|j\ket \bra 0|\right)\nonumber\\
&=& \frac{1}{L} \sum_{j,k\in A}e^{ip(j-k)}|j\ket \bra k| +\frac{L-\ell}{L}  |0\ket \bra 0|,
\eeqa
where $|0\ket$ represents the vacuum state. To compute the R\'enyi entropy we need to take the $n$-th power of the matrix above and then the trace thereof. When doing so the two contributions do not mix so we can write:
\beq
\mathrm{Tr}_A\rho_A^n=\mathrm{Tr}_A (\frac{1}{L} \sum_{j,k\in A} e^{i p(j-k)}|j\ket \bra k|)^n +\mathrm{Tr}_A(\frac{L-\ell}{L}  |0\ket \bra 0|)^n
\eeq
The first term is
\beqa 
\mathrm{Tr}_A (\frac{1}{L} \sum_{j,k\in A} e^{ip(j-k)}|j\ket \bra k|)^n&=& \mathrm{Tr}_A \left(\frac{1}{L^n} \sum_{{k_1,\ldots, k_n \in A}\atop {j_1, \ldots, j_n\in A}} e^{ip j_1}|j_1\ket \left[ \prod_{i=1}^{n-1}e^{ip(j_{i+1}-k_i)}\bra k_i|j_{i+1}\ket\right]  e^{-i p k_n}\bra k_n|\right)\nonumber\\
& = & \mathrm{Tr}_A \left(\frac{1}{L^n} \sum_{{k_1,\ldots, k_n \in A}\atop {j_1, \ldots, j_n\in A}}e^{ip j_1}|j_1\ket \left[ \prod_{i=1}^{n-1}\prod_{i=1}^{n-1}e^{ip(j_{i+1}-k_i)}\delta_{k_i j_{i+1}}\right]  e^{-ip k_n}\bra k_n|\right)\nonumber\\
&=& \frac{1}{L^n} \sum_{{k_1,\ldots, k_n \in A}\atop {j_1, \ldots, j_n\in A}}\delta_{j_1k_n} \left[ \prod_{i=1}^{n-1}\delta_{k_i j_{i+1}}\right] =\frac{\ell^n}{L^n}=r^n.
\eeqa 
The second term is simply
\beq
\mathrm{Tr}_A(\frac{L-\ell}{L}  |0\ket \bra 0|)^n=\frac{(L-\ell)^n}{L^n}=(1-r)^n.
\eeq
Which gives the known formula for the entanglement of a single excitation:
\beq 
S_n^1(r)=\frac{\log(r^n+(1-r)^n)}{1-n}.
\label{fun1}
\eeq 
Note that if the ground state is factorizable (as in the XXZ gapped chain) and therefore has zero entropy  the result gives us the entanglement of the excited state directly. Note that all functions (\ref{fun1}) have a maximum for $r=\frac{1}{2}$ 
\beq
S_n^1(1/2)=\log 2,
\eeq
including the limits $n\rightarrow 1$ and $n\rightarrow \infty$ (see FIG.~1).
The fact that one-particle excitations contribute $\log 2$ to the entanglement entropy was studied in detail in \cite{ln2} where it was shown to be case even for
a non-integrable spin chain model. 
\subsection{Two-Magnon State}
The effect of the presence of non-trivial interaction can be explored by considering a two-magnon state such as
(\ref{two}). 
\subsection{State Normalization}
We will start by fixing the normalization of the state, $N$. The norm is
\beqa
{}_2\bra \Psi|\Psi\ket_2&=& \frac{1}{N}  \sum_{{j_1,j_2, j_1' j_2'=1}}^LS_{j_1 j_2} S_{j_1' j_2'}^* e^{i p_1 (j_1-j_1')+ip_2 (j_2-j_2')} \bra j_1' j_2'|j_1 j_2\ket\nonumber\\
&=& \frac{1}{N}  \sum_{{j_1,j_2, j_1' j_2'=1}}^LS_{j_1 j_2} S_{j_1' j_2'}^* e^{i p_1 (j_1-j_1')+ip_2 (j_2-j_2')} [\delta_{j_1 j_1'}\delta_{j_2 j_2'}+\delta_{j_1 j_2'} \delta_{j_2 j_1'}]\nonumber\\
&=&\frac{1}{N} \sum_{j_1, j_2=1}^L\left[1-\delta_{j_1j_2}+ S_{j_1 j_2} S_{j_2 j_1}^*  e^{i p_{12}j_{12}}\right],
\eeqa
where $p_{12}:=p_1-p_2$ and $j_{12}:=j_1-j_2$. Using the definition of the $S$-matrix we have that we can rewrite the sum as
\beqa
{}_2\bra \Psi|\Psi\ket_2&=&\frac{L(L-1)}{N}+\frac{1}{N} \sum_{j_1=2}^L \sum_{j_2=1}^{j_1-1} \left[S_{j_1 j_2} S_{j_2 j_1}^*  e^{i p_{12}j_{12}}+S_{j_2 j_1} S_{j_1 j_2}^*  e^{i p_{12}j_{21}}\right]\nonumber\\
&=&\frac{L(L-1)}{N}+\frac{1}{N} \sum_{j_1=2}^L \sum_{j_2=1}^{j_1-1} \left[e^{i\varphi}  e^{i p_{12}j_{12}}+e^{-i\varphi}  e^{i p_{12}j_{21}}\right]\nonumber\\
&=&\frac{L(L-1)}{N}+\frac{1}{N} \sum_{j_1=2}^L \sum_{j_2=1}^{j_1-1} \left[2\cos(\varphi+  p_{12}j_{12})\right]\nonumber\\
&=&\frac{L(L-1)}{N}+\frac{1}{N} \frac{(L-1)\cos\varphi-L \cos(\varphi+p_{12})+\cos(\varphi+L p_{12})}{\cos p_{12}-1}. 
\eeqa 
Thus,
\beq
N={L(L-1)}+ \frac{(L-1)\cos\varphi-L \cos(\varphi+p_{12})+\cos(\varphi+L p_{12})}{\cos p_{12}-1}.
\eeq
This same formula was given also in one of the appendices in \cite{Vincenzo}. 
Clearly, for $L$ large and $p_{12}\neq 0$, the second contribution is sub-leading so that 
\beq
N\approx L^2 \quad \mathrm{for} \quad L \quad \mathrm{large}.
\label{nlarge}
\eeq
\subsection{Computation of the Reduced Density Matrix}
We now need to proceed as for the first example. First we compute the reduce density matrix
\beqa
\rho_A=\frac{1}{N} \mathrm{Tr}_B \left(\sum_{j_1,j_2,k_1,k_2=1}^L S_{j_1j_2}S_{k_1 k_2}^*e^{ip_1 (j_1-k_1)+ip_2(j_2-k_2)} |j_1 j_2\ket \bra k_1 k_2| \right).
\eeqa
The only non-vanishing contributions to this trace come from those terms were either all indices are in $A$, all are in $B$ or two indices are in $A$ and two indices are in $B$. This gives the following six contributions:
\beqa
\rho_A&=&\frac{1}{N}\sum_{j_1,j_2,k_1,k_2 \in A} S_{j_1j_2}S_{k_1 k_2}^*e^{ip_1 (j_1-k_1)+ip_2(j_2-k_2)} |j_1 j_2\ket \bra k_1 k_2| \, \mathrm{Tr}_B(|0\ket \bra 0|)\nonumber\\
&+& \frac{1}{N}\sum_{j_1,j_2,k_1,k_2 \in B} S_{j_1j_2}S_{k_1 k_2}^*e^{ip_1 (j_1-k_1)+ip_2(j_2-k_2)} |0\ket \bra 0| \, \mathrm{Tr}_B\left( |j_1 j_2\ket \bra k_1 k_2| \right)\nonumber\\
&+&  \frac{1}{N}\sum_{j_1,k_1 \in A, j_2, k_2 \in B} S_{j_1j_2}S_{k_1 k_2}^*e^{ip_1 (j_1-k_1)+ip_2(j_2-k_2)} |j_1\ket \bra k_1| \, \mathrm{Tr}_B\left( |j_2\ket \bra  k_2| \right)\nonumber\\
&+&  \frac{1}{N}\sum_{j_1,k_2 \in A, j_2, k_1 \in B} S_{j_1j_2}S_{k_1 k_2}^*e^{ip_1 (j_1-k_1)+ip_2(j_2-k_2)} |j_1\ket \bra k_2| \, \mathrm{Tr}_B\left( |j_2\ket \bra  k_1| \right)\nonumber\\
&+&  \frac{1}{N}\sum_{j_2,k_1 \in A, j_1, k_2 \in B} S_{j_1j_2}S_{k_1 k_2}^*e^{ip_1 (j_1-k_1)+ip_2(j_2-k_2)} |j_2\ket \bra k_1| \, \mathrm{Tr}_B\left( |j_1\ket \bra  k_2| \right)\nonumber\\
&+&  \frac{1}{N}\sum_{j_2,k_2 \in A, j_1, k_1 \in B} S_{j_1j_2}S_{k_1 k_2}^*e^{ip_1 (j_1-k_1)+ip_2(j_2-k_2)} |j_2\ket \bra k_2| \, \mathrm{Tr}_B\left( |j_1\ket \bra  k_1| \right)
\eeqa
Noting that
\beq
\mathrm{Tr}_B(|j_1\ket \bra k_1|)=\delta_{j_1 k_1},\quad \mathrm{Tr}_B(|j_1 j_2\ket \bra k_1 k_2|)=\delta_{j_1 k_1}\delta_{j_2 k_2}+\delta_{j_1 k_2}\delta_{j_2k_1},
\eeq 
and grouping all four last terms together by relabelling the summation indices,
the reduced density matrix simplifies to:
\beqa
\rho_A&=&\frac{1}{N}\sum_{j_1,j_2,k_1,k_2 \in A} S_{j_1j_2}S_{k_1 k_2}^*e^{ip_1 (j_1-k_1)+ip_2(j_2-k_2)} |j_1 j_2\ket \bra k_1 k_2| \nonumber\\
&+& \frac{1}{N}\sum_{j_1,j_2,k_1,k_2 \in B} S_{j_1j_2}S_{k_1 k_2}^*e^{ip_1 (j_1-k_1)+ip_2(j_2-k_2)} |0\ket \bra 0| \, (\delta_{j_1 k_1}\delta_{j_2 k_2}+\delta_{j_1 k_2}\delta_{j_2k_1})\nonumber\\
&+&  \frac{1}{N}\sum_{j_1,k_1 \in A, j_2, k_2 \in B} (S_{j_1j_2}S_{k_1 k_2}^*e^{ip_1 (j_1-k_1)+ip_2(j_2-k_2)}+S_{j_1j_2}S_{k_2 k_1}^*e^{ip_1 (j_1-k_2)+ip_2(j_2-k_1)}\nonumber\\
&& +S_{j_2j_1}S_{k_1 k_2}^*e^{ip_1 (j_2-k_1)+ip_2(j_1-k_2)}+S_{j_2 j_1}S_{k_2 k_1}^*e^{ip_1 (j_2-k_2)+ip_2(j_1-k_1)}
) |j_1\ket \bra k_1| \,\delta_{j_2 k_2}.
\eeqa
In the last sum all $S$-matrices have one index in region $A$ and the other in region $B$ so we can use the $S$-matrix definition to simplify the expression. Applying the $\delta$-functions we can write,
\beqa
\rho_A=\frac{1}{N}(\rho_A^{(1)}+\rho_A^{(2)}+\rho_A^{(3)}),
\label{roa}
\eeqa
where we introduced the matrices
\beqa
\rho_A^{(1)}&=&\sum_{j_1,j_2,k_1,k_2 \in A} S_{j_1j_2}S_{k_1 k_2}^*e^{ip_1 (j_1-k_1)+ip_2(j_2-k_2)} |j_1 j_2\ket \bra k_1 k_2| \\
\rho_A^{(2)}&=& \sum_{j_1,j_2\in B} (S_{j_1j_2}S_{j_1 j_2}^*+S_{j_1j_2}S_{j_2 j_1}^* e^{i p_{12} j_{12}} )|0\ket \bra 0| \\
\rho_A^{(3)}&=&  \sum_{{{j_2\in B}}\atop {j_1,k_1 \in A}} [e^{ip_1(j_1-k_1)}+e^{i p_2(j_1-k_1)}+e^{ip_1 j_{12}+ip_2(j_2-k_1)-i\varphi}+ e^{ip_2 j_{12}+ip_1(j_2-k_1)+i\varphi} ]|j_1\ket \bra k_1|.
\eeqa
\subsection{2nd R\'enyi Entropy}
To compute the second R\'enyi entropy we must compute the square of the expression above and then take the trace over subspace $A$. It is easy to show that the only contributions to the trace will come from the squaring each of the three terms above separately, so we can ignore cross-terms that will vanish under the trace. 

For the first term squaring and taking the trace gives
\beqa
\mathrm{Tr}_A((\rho_A^{(1)})^2)&=&\sum_{{j'_1,j'_2,k'_1,k'_2\in A}\atop {j_1,j_2,k_1,k_2 \in A}} S_{j_1j_2}S_{k_1 k_2}^*S_{j'_1j'_2}S_{k'_1 k'_2}^*e^{ip_1 (j_1+j'_1-k_1-k'_1)+ip_2(j_2+j'_2-k_2-k_2')} \nonumber\\
&&\qquad \qquad  \times(\delta_{j_1 k'_1}\delta_{j_2 k'_2}+\delta_{j_1 k'_2}\delta_{j_2 k'_1})
(\delta_{k_1 j'_1}\delta_{k_2 j'_2}+\delta_{k_1 j'_2}\delta_{k_2 j'_1}) \nonumber\\
&&= \sum_{{j_1,j_2,k_1,k_2 \in A}} S_{j_1j_2}S_{k_1 k_2}^*e^{ip_1 (j_1-k_1)+ip_2(j_2-k_2)} \nonumber\\
&& \times \left(S_{k_1 k_2}S_{j_1 j_2}^* e^{i p_1(k_1-j_1)+i p_2(k_2-j_2)}+ S_{k_2 k_1}S_{j_1 j_2}^* e^{i p_1(k_2-j_1)+i p_2(k_1-j_2)} \right. \nonumber\\
&& \left. +S_{k_1 k_2}S_{j_2 j_1}^* e^{i p_1(k_1-j_2)+i p_2(k_2-j_1)}+S_{k_2 k_1}S_{j_2 j_1}^* e^{i p_1(k_2-j_2)+i p_2(k_1-j_1)}  \right),
\eeqa
where, in the first equality, we used the fact that
\beq
\mathrm{Tr}_A (|j_1 j_2\ket \bra k_1 k_2|j'_1 j_2'\ket \bra k'_1 k'_2|)=(\delta_{j_1 k'_1}\delta_{j_2 k'_2}+\delta_{j_1 k'_2}\delta_{j_2 k'_1})
(\delta_{k_1 j'_1}\delta_{k_2 j'_2}+\delta_{k_1 j'_2}\delta_{k_2 j'_1}).
\eeq
Employing the explicit form of the $S$-matrices, the expression can be greatly simplified and factorizes as
\beqa
\mathrm{Tr}_A((\rho_A^{(1)})^2)=\left[\sum_{{k_1  k_2 \in A}} \left(1-\delta_{k_1 k_2}+ S_{k_1 k_2}^* S_{k_2 k_1}  e^{i p_{12}k_{21}}\right)\right]^2.
\eeqa
The sum over the $S$-matrices can be computed by for instance splitting it into the contribution with $k_1>k_2$ and the contribution with $k_1< k_2$. For example
\beqa
\sum_{{k_1 k_2 \in A}} S_{k_1 k_2}^* S_{k_2 k_1} e^{i p_{12}k_{21}}&=& \sum_{{k_1 k_2 \in A}} S_{k_1 k_2}^* S_{k_2 k_1} e^{i p_{12}k_{21}}\nonumber\\
&=&\sum_{k_1=1}^\ell \sum_{k_2=1}^{k_1-1}(S_{k_1 k_2}^* S_{k_2 k_1} e^{i p_{12}k_{21}}+S_{k_2 k_1}^* S_{k_1 k_2} e^{i p_{12}k_{12}})\nonumber\\
&=&\sum_{k_1=1}^\ell \sum_{k_2=1}^{k_1-1}(e^{-i\varphi} e^{i p_{12}k_{21}}+e^{i\varphi} e^{i p_{12}k_{12}})\nonumber\\
&=& 2\sum_{k_1=1}^\ell \sum_{k_2=1}^{k_1-1}\cos(\varphi+p_{12}k_{12})\nonumber\\
&=&\frac{(\ell-1)\cos\varphi-\ell\cos(\varphi+p_{12})+\cos(\varphi+\ell p_{12})}{\cos p_{12}-1}.
\eeqa
Thus, the full contribution to the 2nd R\'enyi entropy (up to normalization of the state) is given by
\beqa
\mathrm{Tr}_A((\rho_A^{(1)})^2)=\left[\ell(\ell-1)+ \frac{(\ell-1)\cos\varphi-\ell\cos(\varphi+p_{12})+\cos(\varphi+\ell p_{12})}{\cos p_{12}-1}\right]^2
\eeqa
Clearly, for $L, \ell$ large and $p_{12}\neq 0$ and considering (\ref{nlarge}) the leading contribution to the entanglement entropy is
\beq
\frac{1}{N^2}\mathrm{Tr}_A((\rho_A^{(1)})^2)\approx r^4.
\label{first}
\eeq
It is easy to show that the contribution to the 2nd R\'enyi entropy of the the second term in (\ref{roa}) is identical to the above, with the replacement $\ell\rightarrow L-\ell$. That is,
\beqa
&&\mathrm{Tr}_A((\rho_A^{(2)})^2)=\nonumber\\
&& \left[(L-\ell)(L-\ell-1)+ \frac{(L-\ell-1)\cos\varphi-(L-\ell)\cos(\varphi+p_{12})+\cos(\varphi+(L-\ell) p_{12})}{\cos p_{12}-1}\right]^2.
\eeqa
for $L, \ell$ large and $p_{12}\neq 0$ the leading contribution is
\beq
\frac{1}{N^2}\mathrm{Tr}_A((\rho_A^{(2)})^2)\approx (1-r)^4.
\label{second}
\eeq
Let us now consider then the third term $\rho_A^{(3)}$. After computing the square and then the trace over subspace $A$ and noting that
\beq
\mathrm{Tr}_A(|j_1\ket \bra k_1| j'_1\ket \bra k'_1|)=\delta_{j_1k'_1}\delta_{j'_1 k_1},
\eeq
we find
\beqa
&& \mathrm{Tr}_A((\rho_A^{(3)})^2)=\sum_{{j_2, j'_2 \in B}\atop {j_1,k_1 \in A}} (e^{ip_1(j_1-k_1)}+e^{i p_2(j_1-k_1)}+e^{-i\varphi} e^{ip_1 j_{12}+ip_2(j_2-k_1)}+e^{i\varphi} e^{ip_2 j_{12}+ip_1(j_2-k_1)} )\nonumber\\
 &&\qquad \quad  \times (e^{ip_1(k_1-j_1)}+e^{i p_2(k_1-j_1)}+e^{-i\varphi} e^{ip_1 (k_1-j'_2)+ip_2(j'_2-j_1)}+e^{i\varphi} e^{ip_2 (k_1-j'_2)+ip_1(j'_2-j_1)} )) 
\eeqa
Expanding the product and simplifying we end up with the sum
\beqa
\mathrm{Tr}_A((\rho_A^{(3)})^2)&=&\sum_{{j_2, j'_2 \in B} \atop {j_1,k_1 \in A}}\left[2+2\cos(p_{12}(j_1-k_1))+2\cos(\varphi+p_{12}(j'_2-j_1))\right.
\nonumber\\
&& \left.+2\cos(\varphi+p_{12}(j'_2-k_1))+2\cos(\varphi+p_{12}j_{21})+2\cos(\varphi+p_{12}(j_2-k_1))\right. \nonumber\\
&& \left.+2\cos(2\varphi+p_{12}(j'_2+j_2-j_1-k_1))+2 \cos(p_{12}(j_2-j'_2)) \right]
\eeqa
This is equal to
\beqa
\mathrm{Tr}_A((\rho_A^{(3)})^2)&=&
2\ell^2 (L-\ell)^2+ 2(L-\ell)^2\frac{\cos(\ell p_{12})-1}{\cos p_{12}-1}+2\ell^2\frac{\cos(L-\ell)p_{12}-1}{\cos p_{12}-1}.\nonumber\\
&& +  2\frac{\cos(2\varphi+ L p_{12})\sin^2\frac{\ell p_{12}}{2}\sin^2\frac{(L-\ell) p_{12}}{2}  }{\sin^4\frac{p_{12}}{2}}\nonumber\\
&& +8 \ell (L-\ell) \frac{\cos(\varphi+\frac{Lp_{12}}{2})\sin\frac{\ell p_{12}}{2}\sin\frac{p_{12}(L-\ell)}{2}}{\sin^2\frac{p_{12}}{2}}
\eeqa
For $\ell, L$ large and $p_{12}\neq 0$  the first term in the sum is leading so the leading contribution to the R\'enyi entropy is
\beq
\frac{1}{N^2}  \mathrm{Tr}_A((\rho_A^{(3)})^2) \approx 2 r^2 (1-r)^2.
\label{third}
\eeq
Putting the contributions (\ref{first}), (\ref{second}) and (\ref{third}) together we find that the 2nd R\'enyi entropy of a two-magnon state for large volume and region size is, as expected.
\beq
S_2^{1,1}(r)=-2\log(r^2+(1-r)^2)=2 S_2^1(r),
\eeq
that is, as expected, twice the entropy of a single excitation. Crucially, for large volume, the result is independent of the scattering matrix. 
\subsection{Equal Momenta}
Although in many cases (such as the XXZ chain) the momenta of the magnons is required to be distinct, it is possible to consider spin chains with bosonic statistics, so that $p_{12}=0$ is allowed. In such cases the formulae of the previous section can still be used, but the terms involving trigonometric functions, which were negligible at large volume, are no-longer so. Indeed, they now become of the same order as the leading terms and it is a simple calculation to show that:
\beq
N=L(L-1)(1+\cos\varphi),
\eeq
and
\beq
\lim_{p_{12} \rightarrow 0}  \mathrm{Tr}_A((\rho_A^{(1)})^2)= \ell^2(\ell-1)^2(1+\cos\varphi)^2,
\label{11}
\eeq
\beq
\lim_{p_{12} \rightarrow 0}  \mathrm{Tr}_A((\rho_A^{(2)})^2)= (L-\ell)^2(L-\ell-1)^2(1+\cos\varphi)^2,
\eeq
\beq
\lim_{p_{12} \rightarrow 0}  \mathrm{Tr}_A((\rho_A^{(3)})^2)=4\ell^2 (L-\ell)^2(1+\cos\varphi)^2.
\label{33}
\eeq
Putting these results together, we have that the entanglement entropy is independent of the interaction and this is irrespective of whether or not volume is large. 
For large volume we find that 
\beq
S_2^2(r)=-\log(r^4+4r^2(1-r)^2+(1-r)^4),
\eeq
which is different from the expression for distinct momenta of the previous section. The expression agrees with our findings for the free massive boson QFT  and the harmonic chain, showing that also for discrete systems the entanglement entropy encodes information about the fermionic/bosonic nature of the quasi-particles. From (\ref{11})-(\ref{33}) it is possible to show the the next to leading order correction for equal momenta is:
\beq
S_2^2(r)=-\log(r^4+4r^2(1-r)^2+(1-r)^4)+\frac{1}{L} \left[1-2\frac{r^3+(1-r)^3}{r^4+4r^2(1-r)^2+(1-r)^4}\right]+O(L^{-2})
\eeq
In particular at $r=\frac{1}{2}$ were entanglement is maximal \cite{ln2} this gives the next-to-leading order correction
\beq
S_2^2({1}/{2})=\log \frac{8}{3}+\frac{1}{3L} +O(L^{-2})
\eeq
It would be interesting to derive such higher order corrections from a QFT computation. A discussion of finize-size corrections to the von Neumann entropy of different kinds of excitations in the XXZ spin-$\frac{1}{2}$ chain was also provided in several of the appendices of \cite{Vincenzo}.
\section{Large Volume Corrections to the 2nd R\'enyi Entropy in the Harmonic Chain}
In this paper we have not presented numerical results for the one-dimensional harmonic chain, whose scaling limit is described by a real massive free boson. Our focus has been on higher dimensions as we intent to discuss the one-dimensional case in much detail in a future work \cite{ussoon}. However, it is interesting to present here some data regarding the next-to-leading order corrections to the maximal entanglement of an excitation, as they display similar features as in two dimensions (see FIG.~2). 
\begin{figure}[h!]
 \includegraphics[width=7cm]{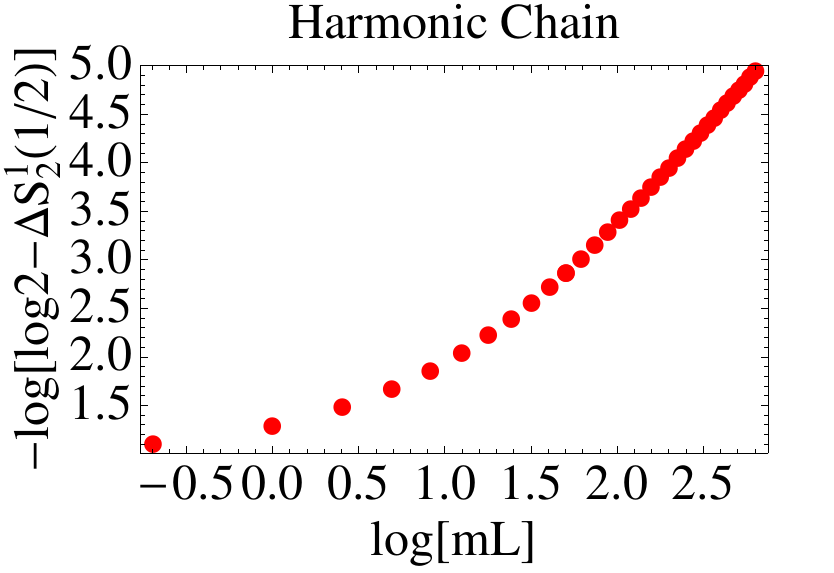} 
 \includegraphics[width=7cm]{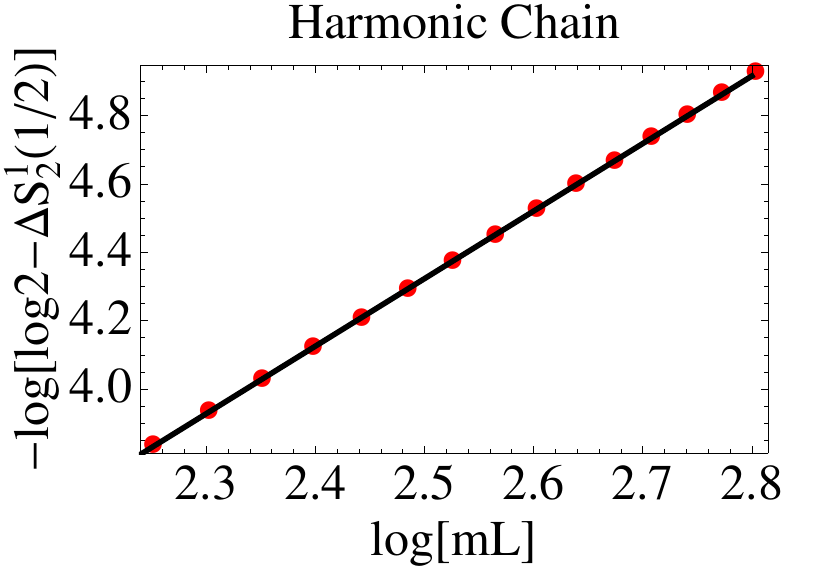} 
  \caption{Left. The figure shows the approach to the value $\log 2$ of the 2nd R\'enyi entropy of a single excitation in a 1D harmonic chain. For large enough volume $mL$
  we see linear behaviour with positive slope, meaning that the leading correction is of the form $\beta/L^\alpha$ for some $\beta$ and $\alpha$. Right. The figure shows a selection of the points on the first figure which are well fitted by the linear fit $-0.613 + 1.974 \log (mL)$ also shown. The fit changes slightly depending on which points are selected. The behaviour strongly suggests (negative) finite-size corrections of order $1/(mL)^2$.}
\end{figure}

An interesting feature of the results above is that the leading corrections to saturation of the entanglement is of order $1/(mL)^2$. From form factor calculations in QFT one would expect a leading correction of order $1/(mL)$ \cite{ussoon} and it would be interesting to investigate whether or not this correction is also vanishing in QFT. The behaviour above is very similar to what we have found for the harmonic lattice. In that case our data are a bit more limited as we do not have access to extremely large surfaces but, as seen in FIG.~\ref{numerics} they also point towards a $1/L^\alpha$  with $\alpha>1$ negative correction to the maximal entanglement of a one-particle state. 

\end{document}